\documentclass[11pt]{article}
\usepackage[centertags]{amsmath}
\usepackage{amssymb,amsfonts}
\usepackage[comma,super]{natbib}
\usepackage{latexsym}
\newcommand{\be}{\begin{equation}} \newcommand{\ee}{\end{equation}}
\newcommand{\bea}{\begin{eqnarray}} \newcommand{\eea}{\end{eqnarray}}
\newcommand{\bse}{\begin{subequations}} \newcommand{\ese}{\end{subequations}}

\begin{document}
\begin{center}
{\Large \bf Generalised isothermal models with strange equation of state} \\
\vspace{1.5cm} {\bf S. D. Maharaj and S. Thirukkanesh$^\dag$}\\
Astrophysics and Cosmology Research Unit,\\
School of Mathematical Sciences,\\
University of KwaZulu-Natal,\\
Private Bag X54001,\\
Durban 4000,\\
South Africa.\\
\vspace{1.5cm} {\bf Abstract}\\
\end{center}
We consider the linear equation of state for matter distributions
that may be applied to strange stars with quark matter. In our
general approach the compact relativistic body allows for
anisotropic pressures in the presence of the electromagnetic field.
New exact solutions are found to the Einstein-Maxwell system. A
particular case is shown to be regular at the stellar centre. In the
isotropic limit we regain the general relativistic isothermal
universe. We show that the mass corresponds to values obtained
previously for quark stars when anisotropy and charge are present.\\
~\\
~\\
{\bf Keywords:} Exact solutions; Einstein-Maxwell spacetimes;
relativistic stars.\\
~\\
{\bf PACS Nos:} 04.20.Jb, 04.40.Nr, 97.10.Cv \\\vspace{5cm}
~\\
$^\dag$Permanent address: Department of Mathematics, Eastern
University, Sri Lanka, Chenkalady, Sri Lanka.

\newpage

\section{Introduction \label{sec:intro}}

In an early and seminal treatment the existence of quark matter in a
stellar configuration in hydrostatic equilibrium was suggested by
Itoh\cite{1}. Subsequently the analysis of strange stars consisting
of quark matter has been considered in a number of investigations.
Strange stars are likely to form in the period of collapse of the
core regions of a massive star after a supernova explosion which was
pointed out by Cheng \emph{et al}\cite{2}. The core of a neutron
star or proto-neutron star is a suitable environment for
conventional barotropic matter to convert into strange quark matter.
Regions of low temperatures and sufficiently high temperatures are
required for a first or second order phase transition which results
in deconfined quark matter. Another possibility suggested by Cheng
and Dai\cite{3} to explain the formation of a strange star is the
accretion of sufficient mass in a rapidly spinning dense star in
X-ray binaries which undergoes a phase transition. The behaviour of
matter at ultrahigh densities for quark matter is not well
understood: in an attempt to study the physics researchers normally
restrict their attention to the MIT bag model (see the treatments of
Chodos \emph{et al}\cite{4}, Farhi and Jaffe\cite{5} and
Witten\cite{6}). The strange matter equation of state is taken to be
\begin{equation}
\label{eq:f} p=\frac{1}{3}(\rho -4B)
\end{equation}
where $\rho$ is the energy density, $p$ is the pressure and $B$ is
the bag constant. The vacuum pressure $B$ is the bag model
equilibrates the pressure and stabilises the system; the constant
$B$ determines the quark confinement. The studies of
Bombaci\cite{7}, Li \emph{et al}\cite{8,9,10}, Dey \emph{et
al}\cite{11}, Xu \emph{et al}\cite{12,13}, Pons \emph{et
al}\cite{14} and Usov\cite{15} directed at particular compact
astronomical objects suggest that these could be strange stars
composed of quark matter with equation of state (\ref{eq:f}).

Mak and Harko\cite{16} found an exact general relativistic model of
a quark star that admits a conformal Killing vector. This was shown
by Komathiraj and Maharaj\cite{17} to be a part of a more general
class of exact analytical models in the presence of the
electromagnetic field with isotropic pressures. The role of
anisotropy was investigated by Lobo\cite{18}, Mak and Harko\cite{19}
and Sharma and Maharaj\cite{20} for strange stars with quark matter
with neutral anisotropic distributions. It is our intention to study
the Einstein-Maxwell system with a linear equation of state with
anisotropic pressures; this treatment would be applicable to a
strange stars which are charged and anisotropic which is the most
general case. In \S2, we write the Einstein-Maxwell system in an
equivalent form using a coordinate transformation. A new exact
solution, in terms of simple elementary functions, is given in \S3.
In addition, we demonstrate that it is possible to find a particular
model which is nonsingular at the stellar origin. The limit of
vanishing anisotropy is studied in \S4 and we regain the isothermal
universes studied previously. In \S5, we consider the physical
features of the new solutions, plot the matter variables for
particular parameter values and show that the quark star mass is
consistent with earlier treatments. Some concluding remarks are made
in \S6.

%
\section{Basic equations \label{sec:fieldeqns}}

It is our intention to model the interior of a dense realistic star
with a general matter distribution. On physical grounds we can take
the gravitational field to be static and spherically symmetric.
Consequently, we assume that the gravitational field of the stellar
interior is represented by the line element
\begin{equation}
\label{eq:f1} ds^{2} = -e^{2\nu(r)} dt^{2} + e^{2\lambda(r)} dr^{2}
+
 r^{2}(d\theta^{2} + \sin^{2}{\theta} d\phi^{2})
\end{equation}
in  Schwarzschild coordinates $(x^{a}) = (t,r,\theta,\phi).$ We
consider the general case of a matter distribution with both
anisotropy and charge. Therefore we take the energy momentum tensor
for the interior to be an  anisotropic charged imperfect fluid; this
is represented by the form
\begin{equation}
\label{eq:f2} T_{ij}=\mbox{diag}\left(-\rho -\frac{1}{2}E^2, p_r-
\frac{1}{2}E^2, p_t+ \frac{1}{2}E^2, p_t+ \frac{1}{2}E^2\right),
\end{equation}
where $\rho$ is the energy density, $p_r$ is the radial pressure,
$p_t$ is the tangential pressure and $E$ is the electric field
intensity. These physical quantities are measured relative to the
comoving fluid velocity $u^i = e^{-\nu}\delta^i_0.$  The line
element (\ref{eq:f1}) and the imperfect matter distribution
(\ref{eq:f2}) generate the Einstein field equations; the field
equations can be written in the form
\begin{eqnarray}
\label{eq:f3} \frac{1}{r^{2}} \left[ r(1-e^{-2\lambda}) \right]' &
=&  \rho + \frac{1}{2}E^{2},\\
 \label{eq:f4} - \frac{1}{r^{2}} \left( 1-e^{-2\lambda} \right) +
\frac{2\nu'}{r}e^{-2\lambda} & = & p_r -\frac{1}{2}E^{2},\\
\label{eq:f5} e^{-2\lambda}\left( \nu'' + \nu'^{2} + \frac{\nu'}{r}-
\nu'\lambda' - \frac{\lambda'}{r} \right)
 & = & p_t + \frac{1}{2}E^{2}, \\
\label{eq:f6} \sigma & = & \frac{1}{r^{2}} e^{-\lambda}(r^{2}E)',
\end{eqnarray}
where primes denote differentiation with respect to $r$ and $\sigma$
is the  proper charge density. We are utilising  units where the
coupling constant $\frac{8\pi G}{c^4}=1$ and the speed of light
$c=1$.  The Einstein-Maxwell system of equations
(\ref{eq:f3})-(\ref{eq:f6}) describes the gravitational behaviour
for an anisotropic charged imperfect fluid. For matter distributions
with $p_r=p_t$ (isotropic pressures) and  $E=0$ (no charge) we
regain Einstein's equations for an uncharged perfect fluid from
(\ref{eq:f3})-(\ref{eq:f6}).

An equivalent form of the field equations is obtained if we
introduce  new variables: the independent variable $x$ and new
functions $y$ and $Z$. These are given by
\begin{equation}
\label{eq:f7} x = Cr^2,~~ Z(x)  = e^{-2\lambda(r)} ~\mbox{and}~
A^{2}y^{2}(x) = e^{2\nu(r)},
\end{equation}
which was earlier used by Durgapal and Bannerji\cite{21} to describe
neutron stars. In terms of the new variables the line element
(\ref{eq:f1}) becomes
\begin{equation}
\label{eq:f8} ds^2 = -A^2 y^2 dt^2 + \frac{1}{4CxZ}dx^2 +
\frac{x}{C} (d\theta^2 +\sin^2\theta d\phi^2),
\end{equation}
where  $A$ and $C$ are arbitrary constants. The transformation
(\ref{eq:f7}) simplifies the field equations, and we find that the
system (\ref{eq:f3})-(\ref{eq:f6}) can be written as
\begin{eqnarray}
\label{eq:f9}\frac{1-Z}{x} - 2\dot{Z} & = & \frac{\rho}{C} +
 \frac{E^{2}}{2C}, \\
\label{eq:f10} 4Z\frac{\dot{y}}{y} + \frac{Z-1}{x} & = &
\frac{p_r}{C}
-  \frac{E^{2}}{2C}, \\
\label{eq:f11} 4x Z \frac{\ddot{y}}{y} +(4 Z+ 2 x \dot{Z})
\frac{\dot{y}}{y} + \dot{Z} &=&\frac{p_t}{C}+ \frac{E^2}{2C}, \\
\label{eq:f12} \frac{\sigma^{2}}{C} & = & \frac{4Z}{x} \left(x
\dot{E} + E \right)^{2},
\end{eqnarray}
where dots denote differentiation with respect to the variable $x$.

The definition
\begin{equation}
\label{eq:f13} m(r)= \frac{1}{2}\int_0^r\omega^2 \rho(\omega)d\omega
\end{equation}
represents the mass contained within a radius $r$ which is a useful
physical quantity. The mass function (\ref{eq:f13}) has the form
\begin{equation}
\label{eq:f14} m(x)=\frac{1}{4 C^{3/2}} \int_0^x\sqrt{w}\rho(w)dw,
\end{equation}
in terms of the new variables introduced  in (\ref{eq:f7}).

On physical grounds we expect that the matter distribution for
realitic stellar matter should satisfy a barotropic equation of
state $p_r=p_r(\rho)$. For the investigations in this paper we
assume the particular equation of state
\begin{equation}
\label{eq:f15} p_r = \alpha \rho - \beta ,
\end{equation}
where $\alpha$ and $\beta$ are constants. This is a simple linear
relationship with desirable physical features and  contains models
investigated previously.  Now it is possible to rewrite
(\ref{eq:f9})-(\ref{eq:f12}) as the system
\begin{eqnarray}
\label{eq:f16}\frac{\rho}{C}&=& \frac{1-Z}{x}-2
\dot{Z}-\frac{E^2}{2C},\\
\label{eq:f17} p_r & =& \alpha \rho -\beta ,\\
\label{eq:f18}p_t &=& p_r +\Delta ,\\
\label{eq:f19} \Delta &=& 4 C x Z \frac{\ddot{y}}{y} + 2 C \left[x
\dot{Z}+ \frac{4Z}{(1+\alpha)}\right]\frac{\dot{y}}{y}+ \frac{(1+5
\alpha)}{(1+\alpha)}C\dot{Z}- \frac{C
(1-Z)}{x}+\frac{2 \beta}{(1+\alpha)},\\
\label{eq:f20}\frac{E^2}{2C}&=& \frac{1-Z}{x}
-\frac{1}{(1+\alpha)}\left[2 \alpha
\dot{Z}+ 4 Z \frac{\dot{y}}{y} +\frac{\beta}{C}\right],\\
\label{eq:f21} \frac{\sigma^2}{C}&=&4 \frac{Z}{x}(x\dot{E}+E)^2,
\end{eqnarray}
where the quantity $\Delta = p_t -p_r$ is defined as the measure of
anisotropy. The Einstein-Maxwell equations as expressed in
(\ref{eq:f16})-(\ref{eq:f21}) is  a system of six nonlinear
equations in terms of eight variables $(\rho, p_r, p_t, \Delta, E,
\sigma,y,Z)$. The system (\ref{eq:f16})-(\ref{eq:f21}) is
under-determined so that there are different ways in which we can
proceed with the integration process. Here we show that it is
possible to specify two of the quantities and generate an ordinary
differential equation in only one dependent variable in the
integration process.  This helps to produce a particular exact
model.
%
%
%
\section{New solutions \label{sec:new solution}}

In this paper, we choose physically reasonable forms for the
gravitational potential $Z$ and electric field intensity $E$ and
then integrate the system (\ref{eq:f16})-(\ref{eq:f21}) to generate
exact models. We make the specific choices
\begin{eqnarray}
\label{eq:f22} Z&=&\frac{1}{a+bx^n},\\
\label{eq:f23}\frac{E^2}{C}&=&\frac{2k(d+ 2x)}{a+bx^n},
\end{eqnarray}
where $a, b, d, n$ and $k$ are real constants. The potential $Z$ is
regular at the origin and continuous in the stellar interior for a
wide range of values for the parameters $a,b$ and $n$. The electric
field intensity $E$ is a bounded and  decreasing function from the
origin to the surface of the sphere. Therefore the forms chosen in
(\ref{eq:f22})-(\ref{eq:f23}) are physically acceptable. These
specific choices for $Z$ and $E$ simplify the integration process.
Equation (\ref{eq:f20}) can be written as
\begin{eqnarray}
\frac{\dot{y}}{y}&=& \frac{(a-1)(1+\alpha)}{4 x}
+\frac{\alpha}{2}\frac{bnx^{n-1}}{(a+b
x^n)}+\frac{(1+\alpha)b}{4}x^{n-1}-\frac{\beta}{4C}(a+bx^n)-
\frac{(1+\alpha)k}{4}(d+2x)\nonumber\\
\label{eq:f24}&&
\end{eqnarray}
where we have used (\ref{eq:f22}) and (\ref{eq:f23}). This has the
advantage of being a first order linear equation in the
gravitational potential $y$.

Equation (\ref{eq:f24}) can be integrated in closed form to give
\begin{equation}
\label{eq:f25} y= D
x^{\frac{(a-1)(1+\alpha)}{4}}(a+bx^n)^{\frac{\alpha}{2}}\exp[F(x)],
\end{equation}
where we have defined
\[F(x)= - \frac{\beta x}{4C}\left[a+\frac{bx^n}{n+1}\right]
+\frac{(1+\alpha)}{4}\left[\frac{bx^n}{n}-k(dx+x^2)\right]\] and $D$
is a constant of integration. Now from (\ref{eq:f22}),
(\ref{eq:f23}) and (\ref{eq:f25}) we can generate an exact model for
the system (\ref{eq:f16})-(\ref{eq:f21}) as follows
\begin{eqnarray}
\label{eq:f26}&& e^{2\lambda}= a+b x^n,  \\
\label{eq:f27}&& e^{2\nu}= A^2
D^2x^{\frac{(a-1)(1+\alpha)}{2}}(a+bx^n)^{\alpha}\exp[2F(x)],\\
\label{eq:f28}&& \frac{\rho}{C}=\frac{(a-1)+bx^n}{x(a+bx^n)}
+\frac{2bnx^{n-1}}{(a+bx^n)^2}-\frac{k(d+2x)}{(a+bx^n)}, \\
\label{eq:f29}&& p_r = \alpha \rho -\beta , \\
\label{eq:f30}&& p_t = p_r+\Delta ,\\
\label{eq:f31}&& \Delta = \nonumber \\
&& \frac{1}{4} \left\{\frac{4C(1-a-bx^n)}{x(a+bx^n)}
-\frac{4bCn(1+5\alpha)x^{n-1}}{(1+\alpha)(a+bx^n)^2}+\frac{8\beta}{(1+\alpha)}
-\frac{2[4a-bx^n(n(1+\alpha)-4)]^2}{(1+\alpha)x(a+bx^n)^3}\times \right.  \nonumber\\
&&\left(a^2(\beta x-C(1+\alpha))+bx^n \left(C\left[(1+\alpha)\left.
\left. (1-bx^n+dkx+2kx^2)-2n \alpha \right]
 +b\beta x^{n+1}\right)
\right. \right. \right. \nonumber\\
 &&\left.   +a [C(1+\alpha)(1+dkx+2kx^2-2bx^n)+2b\beta x^{n+1}]\right)
  +\frac{Cx}{(a+b x^n)}\left[\frac{1}{x^2}\left((a-1)(1+\alpha)\times\right.\right.\nonumber\\
 && \left. \left(\frac{4b n \alpha x^n}{(a+b x^n)}+(a(1+\alpha)-5-\alpha)\right)
  +\frac{4bn \alpha x^n(2a(n-1)+(n \alpha -2)b x^n)}{(a+b x^n)^2}\right) \nonumber\\
 && - \frac{2[a(1+\alpha)(a-1+bx^n)+bx^n(2n\alpha -(1+\alpha))]}{Cx^2(a+b x^n)}
 \left(C(1+\alpha)(k(dx+2x^2)-bx^n)\right)  \nonumber\\
 &&\left. +\beta x(a+b x^n)\right) +4 \left((1+\alpha)
 (b(n-1)x^{n-2}-2k)-\frac{bn \beta x^{n-1}}{C}+\right.  \nonumber\\
&& \left.\left.\left.\frac{[C(1+\alpha)(k(d+2x)-bx^{n-1})
 +\beta (a+bx^n)]^2}{4C^2}\right)\right]\right\}\\
 \label{eq:f32}&&\frac{E^2}{C}=\frac{2k(d+ 2x)}{a+bx^n}.
\end{eqnarray}
The equations (\ref{eq:f26})-(\ref{eq:f32}) represent an exact
solution to the Einstein-Maxwell system
(\ref{eq:f16})-(\ref{eq:f21}) for a charged imperfect fluid with the
linear equation of state $p_r=\alpha \rho -\beta$. Exact solutions
with the equation of state $p_r=\alpha \rho -\beta$ have been used
to model compact objects such as strange stars, as shown by Sharma
and Maharaj\cite{20}, and dark energy stars which are stable, as
demonstrated by Lobo\cite{18}. The exact solution
(\ref{eq:f26})-(\ref{eq:f32}) may be regarded as a generalisation of
an isothermal universe model as we indicate in \S 4.

The solution (\ref{eq:f26})-(\ref{eq:f32}) admits singularities at
the stellar centre in general. The singularity may be avoided for
particular parameter values. If we set $a=1$ and $n=1$ then we
generate the line element
\begin{equation}
\label{eq:f33} ds^2 =-A^2D^2 (1+ bx)^{\alpha} \exp[2F(x)] dt^2+
(1+bx)dr^2 +r^{2}(d\theta^{2} + \sin^{2}{\theta} d\phi^{2}),\nonumber\\
\end{equation}
where $F(x)= -\frac{\beta
x}{8C}(2+bx)+\frac{(1+\alpha)}{4}[bx-k(dx+x^2)]$, with energy
density
\begin{equation}
\label{eq:f34} \frac{\rho}{C}=
\frac{b(3+bx)}{(1+bx)^2}-\frac{k(d+2x)}{(1+bx)}.
\end{equation}
When $k=0$ then the mass function (\ref{eq:f14}) has the form
\begin{equation}
\label{eq:f35} m(x)=\frac{bx^{3/2}}{2C^{3/2}(1+bx)}.
\end{equation}
The expression of the mass function given in (\ref{eq:f35})
represents an energy density which is monotonically decreasing in
the interior of the uncharged sphere and has a finite value at the
centre $x=0$. This  is physically reasonable and similar mass
profiles appear in the treatments of general relativistic
equilibrium configurations of Matese and Whitman\cite{22}, neutron
star models of Finch and Skea\cite{23}, anisotropic stellar
solutions of Mak and Harko\cite{24}, and the dark energy stars of
Lobo\cite{18}. The gravitational potentials for charged imperfect
fluid solution corresponding to the line element (\ref{eq:f33}) are
nonsingular at the origin. The energy density $\rho$, the measure of
anisotropy $\Delta$, and associated quantities $p_r$ and $p_t$ are
also regular at the centre for our choice of $E$. Note that the
quantity $\Delta$ vanishes at the centre and is a continuous
function in the stellar interior. Hence the model
(\ref{eq:f26})-(\ref{eq:f32}) admits a particular case corresponding
to the parameter values $a=1$ and $n=1$ which is regular and well
behaved at the origin.
%
%
\section{Isotropic models \label{sec:iso}}

It is possible to consider the special case of isotropic pressures
with $p_r =p_t$ in the uncharged limit for neutral matter. When
$k=0~(E=0)$ equation (\ref{eq:f31}) becomes
\begin{eqnarray}
\label{eq:f36}&&\Delta =\nonumber\\
&&\frac{1}{4} \left\{\frac{4C(1-a-bx^n)}{x(a+bx^n)}-
\frac{4bCn(1+5\alpha)x^{n-1}}{(1+\alpha)(a+bx^n)}\right.+\frac{8\beta}{(1+\alpha)}
-\frac{2[4a-(n(1+\alpha)-4)bx^n]}{(1+\alpha)x(a+bx^n)^3}\left[a^2\times\right. \nonumber\\
&&\left.\left(\beta x\right.-C(1+\alpha)\right)-bx^n
\left(C\left((1+\alpha)(bx^n-1)\right.\right. \left.+2n\alpha
\right) \left.-b\beta
x^{n+1}\right)+a\left(C(1+\alpha)(1-2bx^2)\right.\nonumber\\
&&\left. \left. +2b\beta
x^{n+1}\right)\right]+\frac{C}{x(a+bx^n)}\left[4b(n-1)(1+\alpha)x^n
\right.+(a-1)(1+\alpha)(a(1+\alpha)-5-\alpha)\nonumber\\
&&+ \frac{4(a-1)bn\alpha(1+\alpha)x^n}{(a+bx^n)}
-\frac{4bn\beta x^{n+1}}{C}+\frac{4bn\alpha [2a(n-1)+b(n \alpha -2)x^n]x^n}{(a+bx^n)^2}\nonumber\\
&&+\frac{[a\beta x-b(C(1+\alpha)-\beta x)x^n]^2}{C^2}+\frac{2}{C(a+b
x^n)}\left(a(1+\alpha)(a-1+bx^n)\right.\nonumber\\
&&\left.+\left(2n\alpha-(1+\alpha)\right)bx^n\right) \left.\left.
\left((C(1+\alpha)-\beta x)bx^n- a\beta x\right)\right]\right\}.
\end{eqnarray}
Equation (\ref{eq:f36}) shows that the model remains anisotropic
even for the uncharged case in general. However for particular
parameter values we can show that $\Delta =0$ in the relevant limit
in the general solution (\ref{eq:f26})-(\ref{eq:f32}). If we set
$b=0$ and $\beta=0$  then (\ref{eq:f36}) becomes
\[\Delta =\frac{C(a-1)}{4ax}\left\{a(1+\alpha)^2-
\left[4 \alpha +(1+\alpha)^2\right]\right\}.\] From the above
equation we easily observe that when $a=1$ and
$a=1+\frac{4\alpha}{(1+\alpha)^2}$  the measure of anisotropy
$\Delta$ vanishes. When $a=1$ we note from (\ref{eq:f28}) that $\rho
=0$ since $b=k=0$. Consequently we cannot regain an isotropic model
when $a=1$; to avoid vanishing energy densities we must have $a\neq
1$ when $\Delta =0$. When $a=1+\frac{4\alpha}{(1+\alpha)^2}$ we
obtain the expressions
\begin{eqnarray}
\label{eq:f37} ds^2 &=&-B r^{\frac{4\alpha}{1+\alpha}} dt^2+
\left[1+\frac{4\alpha}{(1+\alpha)^2}\right]dr^2+r^{2}(d\theta^{2} +
\sin^{2}{\theta} d\phi^{2}), \\
\label{eq:f38} \rho&=& \frac{4 \alpha}{[4 \alpha
+(1+\alpha)^2]r^2},~~ ~p_r(=p_t)=\alpha \rho,
\end{eqnarray}
where we have set $A^2D^2a^{\alpha}=B$ and $C=1$.

The above solution was obtained by Saslaw \emph{et al}\cite{25} in
their investigation of general relativistic isothermal universes.
Since $\rho \propto r^{-2}$ we may interpret (\ref{eq:f37}) as a
relativistic cosmological metric since there is an analogy that can
be made with the well known Newtonian solution as pointed out by
Chandrasekhar\cite{26}. In the Newtonian case the density $\rho$ is
finite in the core regions and decreases according to $r^{-2}$ in
the rest of the model. The total mass and region of the isothermal
model are infinite. Therefore the line element (\ref{eq:f37}) may be
interpreted as a relativistic inhomogeneous universe where the
nonzero pressure balances gravity. Saslaw \emph{et al}\cite{25}
point out that (\ref{eq:f37}) may be viewed as the asymptotic state
of the Einstein-de Sitter cosmological model, in an expansion-free
state as $t \rightarrow \infty$, where the hierarchial distribution
of matter has clustered over large scales. As $t \rightarrow \infty$
the Einstein-de Sitter model tends to the line element
(\ref{eq:f37}) and the universe evolves into an isothermal static
sphere given by the exact solution (\ref{eq:f37}) with equation of
state $p_r(=p_t)=\alpha \rho$. To move from the pressure-free
Einstein-de Sitter model to the static isothermal metric requires a
phase transition for condensation through clustering of galaxies. It
is interesting to observe that Chaisi and Maharaj\cite{27} and
Maharaj and Chaisi\cite{28} have obtained generalised anisotropic
static isothermal spheres by utilising a known isotropic metric to
produce a new anisotropic solution of the Einstein field equations.
Govender and Govinder\cite{29} have found simple nonstatic
generalisations of isothermal universes which describe an isothermal
sphere of galaxies in quasi-hydrostatic equilibrium with heat
dissipation driving the system to equilibrium. Hence the imperfect
relativistic fluid, in the presence of electromagnetic field, with
the linear equation of state $p_r=\alpha \rho - \beta$, is a
generalisation of the conventional isothermal model with a clear
physical basis.
%
%
\section{Physical analysis \label{sec:Schwarz}}

We observe that the exact solution (\ref{eq:f26})-(\ref{eq:f32}) may
be singular at the origin in general. The solution should be used to
describe the gravitational field of the envelope in the outer
regions of a quark star or a dark energy star with equation of state
$p_r=\alpha \rho - \beta$. To avoid the singularity at the centre
another solution is required to model the stellar core. Examples of
core-envelope models in general relativity are provided by Thomas
\emph{et al}\cite{30}, Tikekar and Thomas\cite{31} and Paul and
Tikekar\cite{32}. We observe that if $a=1$ then the imperfect
charged solution (\ref{eq:f26})-(\ref{eq:f32}) admits the line
element
\begin{equation}
\label{eq:f39}ds^2=- A^2D^2(1+b x^n)^{\alpha} \exp[2F(x)]
dt^2+(1+bx^n)dr^2+r^{2}(d\theta^{2} + \sin^{2}{\theta} d\phi^{2}),
\end{equation}
where $F(x)= - \frac{\beta x}{4C}\left[1+\frac{bx^n}{n+1}\right]
+\frac{(1+\alpha)}{4}\left[\frac{bx^n}{n}-k(dx+x^2)\right]$. It is
clear from (\ref{eq:f39}) that the gravitational potentials are
nonsingular at the origin for all values of $n$. If $n=1$ then there
is no singularity in the energy density $\rho$ and the model is
nonsingular throughout the stellar interior. However if $n \neq 1$
then the gravitational potential potentials (\ref{eq:f39}) may
continue to be well behaved but singularities may appear in the
matter variables at the origin $r=0$.

\begin{figure}[thb]
\vspace{1.8in} \includegraphics{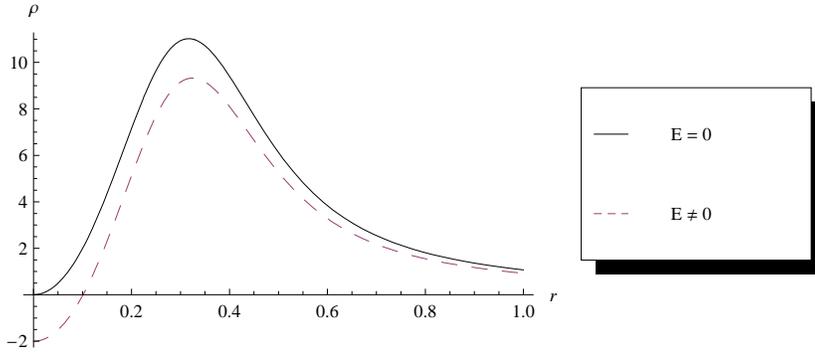} \caption{\label{Gr-density}
Energy density.}
\end{figure}

\begin{figure}[thb]
\vspace{1.8in} \includegraphics{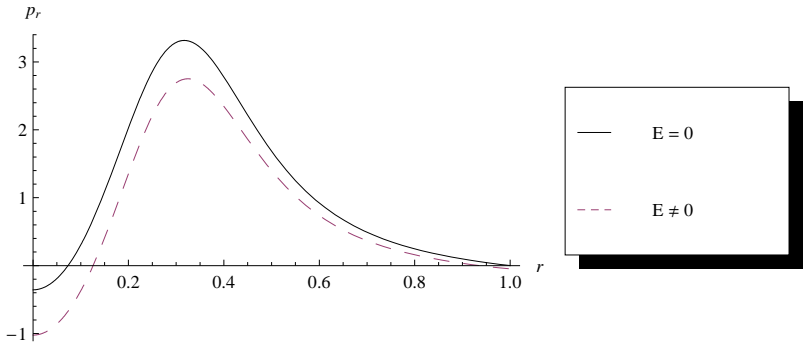}
\caption{\label{Gr-pressurer} Radial pressure.}
\end{figure}

\begin{figure}[thb]
\vspace{1.8in} \includegraphics{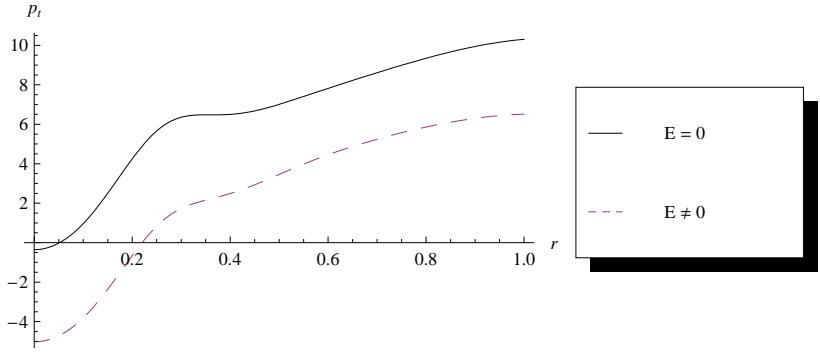}
\caption{\label{gr-tangen} Tangential pressure.}
\end{figure}

\begin{figure}[thb]
\vspace{1.8in} \includegraphics{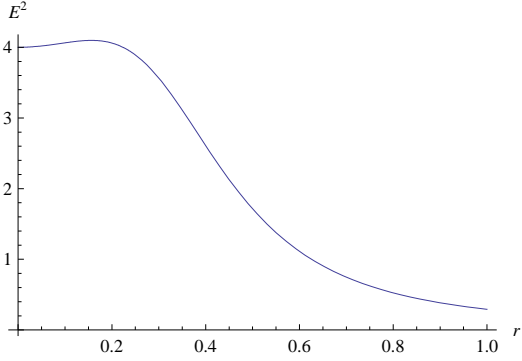}
\caption{\label{Gr-Electricfield} Electric field intensity.}
\end{figure}

\begin{figure}[thb]
\vspace{1.8in} \includegraphics{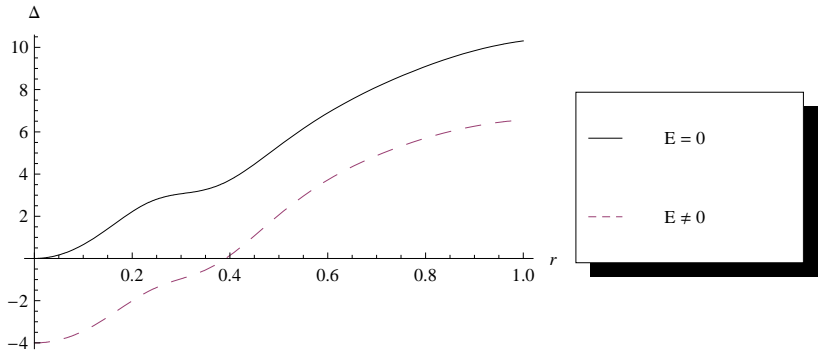} \caption{\label{Gr-del-2}
Measure of anisotropy.}
\end{figure}

By considering a particular example we can demonstrate graphically
that the matter variables are well behaved outside the origin.
Figures \ref{Gr-density}-\ref{Gr-del-2} represent the energy
density, the radial pressure, the tangential pressure, the electric
field intensity and the measure of anisotropy, respectively. Note
that solid lines represent uncharged matter $(E=0)$ and dashed lines
include the effect of charged matter $E\neq 0$. To plot the graphs
we choose the parameters $n=2, a=1, b=40, k=2, d=1, \alpha
=\frac{1}{3}$ and $\beta =0.3569$, and the stellar boundary is set
at $r=1$. From Figures \ref{Gr-density}-\ref{Gr-pressurer} we see
that both the energy density $\rho$ and the radial pressure $p_r$
are continuous throughout the interior, increasing from the centre
to $r=0.32$ and then decreasing.  Note that the radial pressure is
zero at the boundary $r=1$ for the uncharged case $E=0$. We observe
from Figure \ref{gr-tangen} that the tangential press $p_t$ is
continuous and well behaved in the interior regions. From Figure
\ref{Gr-Electricfield} we observe that the electric field intensity
$E$ is decreasing smoothly throughout the stellar interior. We can
observe from Figure \ref{Gr-del-2} that the measure of anisotropy is
continuous throughout the stellar interior. The behaviour of
$\Delta$ outside the centre is likely to correspond to physically
realistic matter in the presence of the electromagnetic field.
Figure 5 has a profile similar to the anisotropic boson stars
studied by  Dev and Gleiser\cite{33} and the compact anisotropic
relativistic spheres of  Chaisi and Maharaj\cite{34}. From the
figures we can see that the effect of electric field intensity $E$
is to produce  lower values for $\rho, p_r, p_t$ and $\Delta$.

\vspace{.25in}

\vspace{.25in}

\vspace{.25in}

\vspace{.25in}

We now show that the solutions generated in this paper can be used
to describe realistic compact objects for the case $n=1$ as an
example from \S3. In our model, when $n=1$, the parameters $b$ has
the dimension of length$^{-2}$, $k$ has the dimension of
length$^{-4}$ and $d$ has the dimension of length$^2$. For
simplicity, we introduce the transformations
\[\tilde{b}=bR^2,~\tilde{k}=kR^4,~\tilde{d}=d R^{-2}, \]
where $R$ is a parameter which has the dimension of a length. Under
these transformations the energy density becomes
\begin{equation}
\rho = \frac{1}{R^2}\left[\frac{\tilde{b}}{1+\tilde{b}y}
+\frac{2\tilde{b}}{(1+\tilde{b}y)^2}-\frac{\tilde{k}(\tilde{d}+2y)}{1+\tilde{b}y}\right],\nonumber
\end{equation}
where we have set $C=1$ and $y= \frac{r^2}{R^2}$. Then  the mass
contained within a radius $s$ has the form
\[M= \frac{1}{2}\left\{\frac{\tilde{b}\frac{s^3}{R^2}}
{1+\tilde{b}\frac{s^2}{R^2}} \frac{2\tilde{k}}{3\tilde{b}}
\frac{s^3}{R^2}+\tilde{k}(2-\tilde{b}\tilde{d})
\left[\frac{s}{\tilde{b}^2}-\frac{R}{\tilde{b}^{5/2}} \arctan
\left[\sqrt{\tilde{b}}\frac{s}{R}\right]\right]\right\}.
\]
For simplicity we set $\tilde{b}\tilde{d}=2$ so that these
expressions reduce to
\begin{eqnarray}
\label{eq:f40}\rho & =&
\frac{1}{R^2}\left[\frac{\tilde{b}}{1+\tilde{b}y}
+\frac{2\tilde{b}}{(1+\tilde{b}y)^2}-\frac{2
\tilde{k}}{\tilde{b}}\right],\\
\label{eq:f41}M &=&
\frac{1}{2}\left[\frac{\tilde{b}}{1+\tilde{b}\frac{s^2}{R^2}}
-\frac{2\tilde{k}}{3\tilde{b}}\right]\frac{s^3}{R^2},
\end{eqnarray}
which are simple forms. It is now easy to calculate the density and
mass for particular parameter values from (\ref{eq:f40}) and
(\ref{eq:f41}). For example, when $s=7.07$km, $R=1$km,
$\tilde{b}=0.03$ and $\tilde{k}=0.00045$, we obtain the mass
$M_{E=0}= 1.436M_{\odot}$ for uncharged matter  $M_{E\neq 0}=
0.240M_{\odot}$ for charged matter. Note that the value of the mass
for uncharged matter is consistent with the strange star models
previously found by Sharma and Maharaj\cite{20} and Dey \emph{et
al}\cite{11}. We have shown that the inclusion of the
electromagnetic field affects the value for the mass $M$. It is also
possible to to relate our results to other treatments. If we set
$s=9.46$km, $R=1$km, $\tilde{b}=0.35$ and $\tilde{k}=0.00045$, then
we obtain the mass $M_{E=0}= 3.103M_{\odot}$ for uncharged matter
and $M_{E\neq 0}= 2.858M_{\odot}$ for charged matter.
 These values for the mass are similar to charged quark stellar
 models generated by  Mak and Harko\cite{16} which describe a unique static charged
 configuration of quark matter admitting a one-parameter group of
 conformal symmetries. We have shown that the presence of anisotropy
 and charge in the matter distribution yields masses which are
 consistent with other investigations.

%
%

\section{Discussion}

We have investigated the Einstein-Maxwell system of field equations
with a strange matter equation of state for anisotropic matter
distributions when the electromagnetic field is present. A new class
of exact solutions was found to this system of nonlinear equations.
A particular case is nonsingular at the centre of the star and the
gravitational potentials and matter variables are well behaved. When
the anisotropy vanishes we regain the general relativistic
isothermal universe of Saslaw\cite{25}. This is an extension of the
conventional Newtonian isothermal universe with density $\rho
\propto r^{-2}$. The plots of the matter variables indicate that the
solution may be used to model quark stars, at least in the envelop
if singularities are present at the origin. We calculated the mass
in a special case $(a=1, n=1)$ and showed that this value is
consistent with strange matter distributions of Dey \emph{et
al}\cite{11}, Mak and Harko\cite{16} and Sharma and
Maharaj\cite{20}. Ours is a particular solution of the
Einstein-Maxwell system with the nice feature of containing the
isothermal universe in the isotropic limit. A more comprehensive
study of other possible solutions admitted, with the strange matter
equation of state, is likely to produce other interesting results.

 We now comment on some physical aspects of the solutions found which
are of interest. Firstly, there are a number of free parameters in
the solution which may be determined by imposing boundary
conditions. The solutions may be connected to the Reissner-Nordstrom
exterior spacetime
\begin{equation}
\label{eq:f41}ds^2=-\left(1-\frac{2M}{r}+\frac{Q^2}{r^2}\right)dt^2
+\left(1-\frac{2M}{r}+\frac{Q^2}{r^2}\right)^{-1}dr^2
+r^2\left(d\theta^{2} + \sin^{2}{\theta} d\phi^{2}\right)
\end{equation}
with the interior spacetime (\ref{eq:f39}) across the boundary $r=s$
where $M$ and $Q$ represent the total mass and the charge of the
star, respectively. This gives the conditions
\begin{eqnarray}
1-\frac{2M}{s}+\frac{Q^2}{s^2}&=& A^2D^2\left\{[1+b(Cs^2)^n]^\alpha
\exp[2F(Cs^2)]\right\} \nonumber\\
&& \nonumber\\
\left(1-\frac{2M}{s}+\frac{Q^2}{s^2}\right)^{-1}&=&1+b(Cs^2)^n.
\nonumber
\end{eqnarray}
Therefore the continuity of the metric coefficients across the
boundary $r=s$ is maintained. The number of free parameters $(A, C,
D, b~\mbox{and}~ n)$ ensures that these necessary conditions are
satisfied. Secondly, from the graphs generated for the energy
density and the pressure we observe that these quantities are
negative in regions close to the centre. We believe that this
feature arises because of the strange equation of state that we have
imposed on the model. To avoid negative matter variables we would
need to use another solution to describe the core with our solution
serving as the envelope. This is the situation that arises in the
gravastar model of Mazur and Mottola where the singular core is
replaced by a de Sitter condensate through a phase transition near
the location of the event horizon. This idea has been utilised by
Lobo\cite{18} to demonstrate that dark energy stars are stable.
Thirdly, the example of this paper and other exact solutions found
for the Einstein-Maxwell system suggests that the negativity of the
energy density and the pressure close to the centre may be a generic
feature of charged stellar models. It appears that the existence of
the electric field always makes the energy density negative close to
the centre of the sphere. This suggests the conjecture that a
charged fluid cannot satisfy the energy conditions near the stellar
centre is possible. This will be the topic of a separate
investigation.

\section*{Acknowledgements}
We are grateful to the referee for insightful comments relating to
the physical properties of the models. ST thanks the National
Research Foundation and the University of KwaZulu-Natal for
financial support, and is grateful to Eastern University, Sri Lanka
for study leave. SDM acknowledges that this work is based upon
research supported by the South African Research Chair Initiative of
the Department of Science and Technology and the National Research
Foundation.
%
%

\end{document}